\documentclass[aps,a4paper,twocolumn,pra]{revtex4}
\pdfoutput=1

\usepackage{amsmath}%
\usepackage{amsfonts}%
\usepackage{amssymb}%
\usepackage[cal=cm]{mathalfa}
\usepackage{graphicx}
\usepackage{braket}
\usepackage{sidecap}
\usepackage{color}
\usepackage{bbm}
\usepackage{nicefrac}
\usepackage{float}
\usepackage[dvipsnames]{xcolor}
\usepackage[utf8]{inputenc}
\usepackage[normalem]{ulem}
\usepackage{soul}
\usepackage{nicefrac}
\usepackage{sidecap}
\usepackage[citebordercolor=blue]{hyperref}
\usepackage{textcomp}

\newcommand{\stkout}[1]{\ifmmode\text{\sout{\ensuremath{#1}}}\else\sout{#1}\fi}

\definecolor{brass}{rgb}{0.71, 0.65, 0.26}
\definecolor{cgreen}{rgb}{0.0, 0.42, 0.24}
\definecolor{cadetblue}{rgb}{0.37, 0.62, 0.63}

\begin{document}

\title{Experimental Test of Tight State-Independent \\Preparation Uncertainty Relations for Qubits}
\author{Stephan Sponar$^1$}
\email[]{stephan.sponar@tuwien.ac.at}
\author{Armin Danner$^1$}
\author{Kazuma Obigane$^2$}
\author{Simon Hack$^1$}
\author{Yuji Hasegawa$^{1,2}$}
\email[]{yuji.hasegawa@tuwien.ac.at}
\affiliation{%
$^1$Atominstitut, TU Wien, Stadionallee 2, 1020 Vienna, Austria \\
$^2$Department of Applied Physics, Hokkaido University, Kita-ku, Sapporo 060-8628, Japan
}

\date{\today}

\hyphenpenalty=800\relax
\exhyphenpenalty=800\relax
\sloppy
\setlength{\parindent}{0pt}

\noindent

\begin{abstract}
The well-known Robertson-Schr\"odinger  uncertainty relations miss an irreducible lower bound. This is widely attributed to the lower bound's state-dependence. Therefore, Abbott \emph{et al.} introduced a general approach to derive tight state-independent uncertainty relations for qubit measurements [Mathematics \textbf{4}, 8 (2016)]. The relations are expressed in two measures of uncertainty, which are standard deviation and entropy, both functions of the expectation value. Here, we present a neutron optical test of the tight state-independent preparation uncertainty relations for non-commuting Pauli spin observables with mixed spin states. The final results, obtained in a polarimetric experiment, reproduce the theoretical predictions evidently for arbitrary initial states of variable degree of polarization.
\end{abstract}

\maketitle


\section{Introduction}

The impossibility of assigning definite values to incompatible observables is a fundamental feature of quantum mechanics. It manifests in the impossibility to \emph{prepare} quantum states that simultaneously have precise values of position $Q$ and momentum $P$. This is expressed in the well-known position-momentum uncertainty relation $\Delta Q\Delta P\geq\nicefrac{\hbar}{2}$, which sets a lower bound on the product of standard deviations of the position and momentum observables. The position-momentum uncertainty relation has been generally proven from basic principles of quantum mechanics by Kennard in 1927 \cite{Kennard27}, following Heisenberg's introduction of the \emph{uncertainty principle} illustrated by the famous $\gamma$-ray microscope \emph{Gedankenexperiment} \cite{Heisenberg27}. However, the  $\gamma$-ray microscope sets a lower bound for the product of the \emph{measurement} error and the disturbance in a joint measurement of position $Q$ and momentum $P$ on a single particle. Hence, the position-momentum uncertainty relation in terms of standard deviations quantifies how precise with respect to the observables of interest, a state can be \emph{prepared}, rather than the ability to jointly \emph{measure} them.

In 1929 Robertson generalized the uncertainty relation to arbitrary pairs of incompatible (i.e., non-commuting) observables $A$ and $B$ as  
\begin{equation}\label{eq:Robertson}
\Delta A\Delta B\geq\vert\frac{1}{2i}\langle\psi\vert [A,B]\vert\psi\rangle\vert,
\end{equation}
for any state $\vert\psi\rangle$, where $[A,B]$ represents the commutator $[A,B]=AB-BA$ and the standard deviation of an observable $X$ is defined as $(\Delta X)^2=\langle\psi\vert X^2\vert\psi\rangle-\langle\psi\vert X\vert\psi\rangle^2$ \cite{Robertson29}. However, Robertson's uncertainty relation turned out to follow from a slightly stronger inequality namely the Schr\"odinger uncertainty relation  \cite{Schroedinger30}, given by 
  \begin{eqnarray}\label{eq:Schroedinger}
 (\Delta A)^2(\Delta B)^2\geq&&\vert\langle\psi\vert \{ A, B\}\vert\psi\rangle-\langle\psi\vert A\vert\psi\rangle \langle\psi\vert B\vert\psi\rangle \vert^2\nonumber\\&&+\vert \frac{1}{2i}\langle\psi\vert[ A, B]\vert\psi\rangle\vert^2,
 \end{eqnarray}
where the anticommutator $\{ A, B\}= A B+ B A$ is used. Here, the right-hand side (RHS) of Eq.\,(\ref{eq:Schroedinger}) yields a tighter bound than Eq.\,(\ref{eq:Robertson}), but not necessarily saturated.

Note that Kennard's, Robertson's and Schr\"odinger's uncertainty relations all express a quantitative statement about the measurement statistics for $A$ and $B$ of different ensembles that are obtained separately (many times) on identically-prepared quantum systems; this is the reason why such relations are usually referred to as \emph{preparation uncertainty relations}. They propose fundamental limits on the measurement statistics for any state preparation.

The fact that in the case of \emph{preparation uncertainty relations} the measurements are performed on different ensembles is in total contrast to Heisenberg's original discussion of his \emph{uncertainty principle}, which addresses the inability to jointly (simultaneously or sequentially) measure incompatible observables with arbitrary accuracy, which is described by \textit{measurement uncertainty relations}. Consequently, uncertainty relations have a long history of being misinterpreted as exclusive statements about joint measurements.

In recent years \textit{measurement uncertainty relations}, as originally proposed by Heisenberg \cite{Heisenberg27}, have received renewed attention. New measures and uncertainty relations for error and disturbance have been proposed~\cite{Ozawa03, Busch13}, refined~\cite{Branciard13, Ozawa14}, and experimentally tested in neutronic  \cite{Erhart12, Sulyok13, Demirel16,Sponar17,Demirel19,Demirel20} and photonic  \cite{Steinberg12,Edamatsu13,Kaneda14,Ringbauer14,Ma16,Pan19} systems. However, there continues to be some debate as to the appropriate measure of measurement (in)accuracy and of disturbance \cite{Ozawa03,OzawaPLA03,Hall04,Branciard13,Erhart12,Steinberg12,Sulyok13,Busch13,Busch13PRA,Busch14,Ringbauer14,Kaneda14,Buscemi14,Sulyok15,Ma16,Sponar17,Barchielli18,Pan19}.

This recent interest in measurement uncertainty relations revealed that the well-known Robertson-Schr\"odinger uncertainty relation lacks an irreducible or \emph{state-independent} lower bound of the RHS of Eq.(\ref{eq:Robertson}). Owing to this fact the lower bound on the right-hand side of Eqs.\,(\ref{eq:Robertson}) and (\ref{eq:Schroedinger}) may become zero for certain states, even for non-commuting $A$ and $B$, as is the case for instance for neutron spins. Hence, Deutsch began to seek a theorem of linear algebra in the form  $\mathcal U(A, B,\psi)\ge\mathcal B( A, B)$ - that is a \emph{state-independent} relation - and furthermore suggested to use (Shannon) \emph{entropy} as measure \cite{Deutsch83,Kraus87}. Note that Heisenberg's (Kennard) inequality $\Delta Q\Delta P\geq\nicefrac{\hbar}{2}$ has that form but its generalizations Eqs.\,(\ref{eq:Robertson}) and (\ref{eq:Schroedinger}) do not. Common to all entropic uncertainty relations is the peculiarity of setting bounds on the sum of the entropies of $A$ and $B$ rather than on the product. The most well known formulation of entropic uncertainty relations was given by Maassen and Uffink \cite{Maassen88} in 1988 as
\begin{equation}\label{eq:Maassen}
H(A)+H(B)\geq -2\,{\rm{log}}\,c,
\end{equation}
where $c={\rm{max}}_{i,j}\,\vert\langle a_i\vert b_j\rangle\vert$ is the maximum overlap between the eigenvectors $\vert a_i\rangle$ and $\vert b_j\rangle$ of observables $A$ and $B$, respectively. Then the Shannon entropy $H(A)=\sum_i {\rm{Tr}}[\rho P_i] {\rm{log}} ( {\rm{Tr}}[\rho P_i]) $, with $P_i$ being a projector from the spectral decomposition of the observable $A$, given by $A=\sum_i a_iP_i$, provides a measure of uncertainty for the observable $A$ in the state $\rho$. In more recent time, entropic uncertainty relations have been extended to include the case of quantum memories \cite{Berta2010,Pati12}.


The growth in the numbers of studies, both theoretically and experimentally, in \emph{measurement} uncertainty relations has prompted renewed interest in the possibility of state-independent preparation uncertainty relations for the standard deviations of observables, rather than entropic relations.

\section{Theoretical framework}

In \cite{Abbott16} Abbott and Branciard proposed an approach for deriving tight state-independent and \emph{partially} state-dependent (that is depending on the mixing parameter $r$ of non-pure states) uncertainty relations for qubit measurements that completely characterize the obtainable uncertainty values. Their equivalent relations in terms of expectation values, standard deviations and entropies can more generally be transformed into other relations, in terms of any measure of uncertainty that can be written as a function of the expectation value. Any pair of Pauli observables $A=\vec a\cdot\vec\sigma$ and $B=\vec b\cdot\vec\sigma$, with $\vec \sigma=(\sigma_x,\sigma_y,\sigma_z)^T$  and an arbitrary quantum state $\rho=\frac{1}{2}({1\!\!1}+\vec r\cdot\vec\sigma)$ satisfies the condition 
\begin{eqnarray}\label{eq:ExpVal}
\vert\langle A\rangle\vec a-\langle B\rangle\vec b\vert^2\leq(1-(\vec a\cdot\vec b)^2)\vert \vec r\vert^2\nonumber\\
\leq1 -(\vec a\cdot\vec b)^2=\vert\vec a\times\vec b\vert^2.
\end{eqnarray}
The standard deviation $\Delta A$ and expectation value $\langle  A\rangle$ are connected via  
\begin{eqnarray}
(\Delta  A)^2=1-\langle  A\rangle^2\quad\textrm{and}\quad\langle  A\rangle=\pm\sqrt{1-(\Delta  A)^2},
\end{eqnarray}
since every Pauli operator $A$ satisfies $\langle  A^2\rangle={1\!\!1}$. Hence, the tight state-independent uncertainty relation, given in Eq.(\ref{eq:ExpVal}), can be rewritten in terms of standard deviations as 
\begin{eqnarray}\label{eq:StDec}
(\Delta  A)^2+(\Delta  B)^2+2\vert\vec a\cdot\vec b\vert\sqrt{1-(\Delta  A)^2}\sqrt{1-(\Delta  B)^2}\nonumber\\
\ge2-\big(1-(\vec a\cdot\vec b)^2\big)\vert\vec r\vert^2\ge1+(\vec a\cdot\vec b)^2.
\end{eqnarray}
In the case of qubits, the Shannon entropy of a Pauli observable $H( A)$ can be directly expressed in terms of the expectation value $\langle  A\rangle$, namely: 
\begin{eqnarray}
H( A)=h_2\Bigg(\frac{1+\langle A\rangle}{2}\Bigg)=h_2\Bigg(\frac{1-\langle A\rangle}{2}\Bigg),
\end{eqnarray}
where $h_2$ is the binary entropy function defined as 
\begin{eqnarray}
h_2(p)=-p\,{\mathrm {log}}\,p-(1-p)\,{\mathrm {log}}\,(1-p),
\end{eqnarray}
 or $\langle A\rangle=\pm f(H( A))$ with $f(x):=1-2h^{-1}_2(x)$, where   $h^{-1}_2$ denotes the inverse function of  $h_2$. Then one obtains the following tight relation for two Pauli observables 
\begin{eqnarray}\label{eq:Entropy}
f\big(H(A)\big)^2+f\big(H(B)\big)^2-2\vert\vec a\cdot\vec b\vert\,f\big(H(A)\big)\,f\big(H(B)\big)\nonumber\\
\leq\big(1-(\vec a\cdot\vec b)^2\big)\vert\vec r\vert^2\leq 1-(\vec a\cdot\vec b)^2.
\end{eqnarray}
 Note that the uncertainty relations in terms of the standard deviations and the entropy, which are given by Eqs. (\ref{eq:StDec}) and (\ref{eq:Entropy}), are tight (state-independent) relations.

\begin{SCfigure*}
	\includegraphics[width=0.73\textwidth]{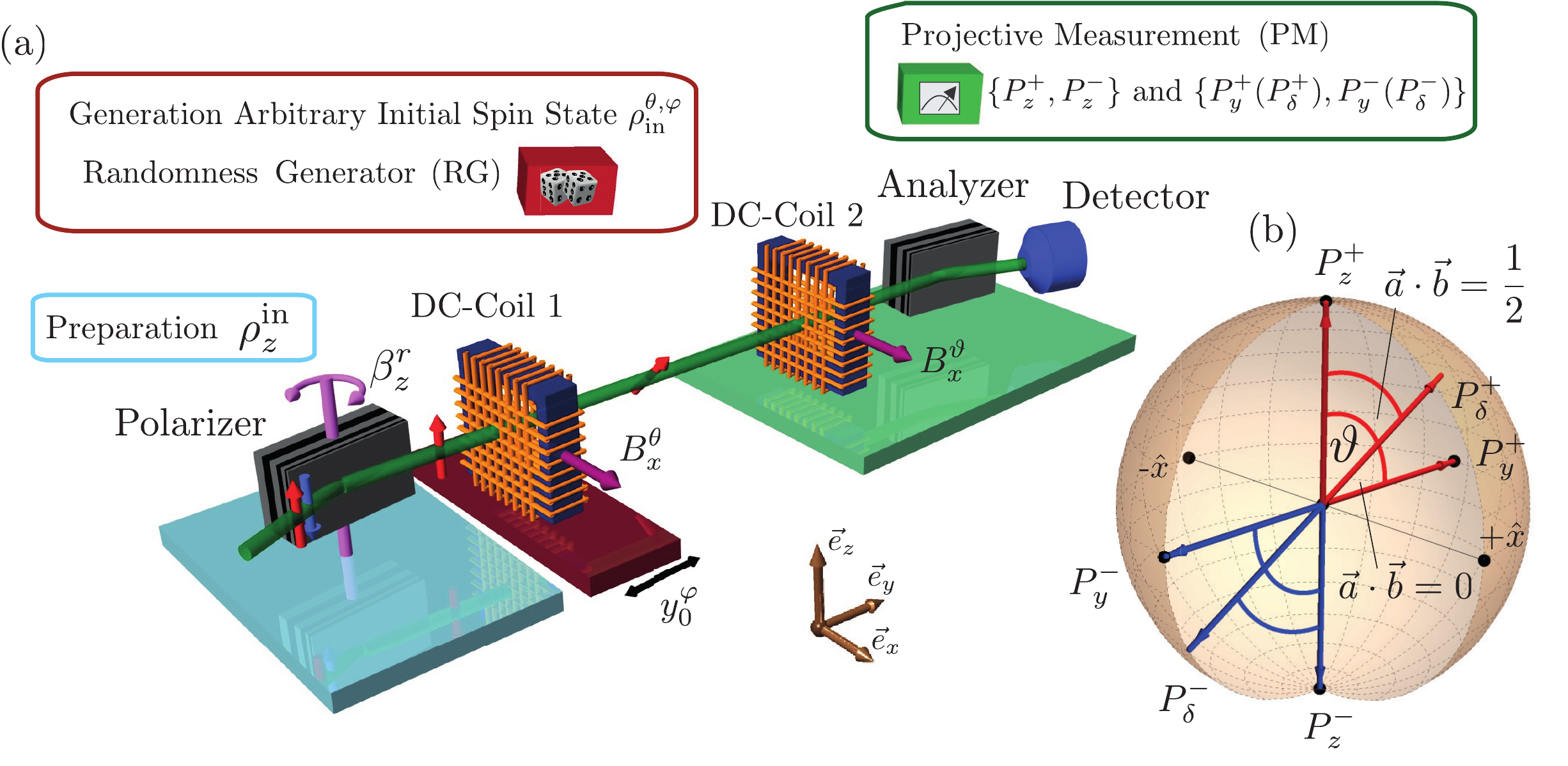}
	\caption{(a) Experimental setup for determining the tight preparation uncertainty relation for neutron spins. Blue: preparation of incident state $\rho_z^{\rm{in}}$. Ruby: the initial state $\rho_{\rm{in}}^{\theta,\varphi}$ is chosen by the randomness generator (RG) and generated by DC-coil 1, where polar angle $\theta$ is dependent on the static magnetic field $B_x^{\rm{RG}}$ and azimuthal angle $\varphi$ on the position $y_0$. Green: the projectors $P^+_z,P^-_z,P^+_y(P^+_\delta)$ and $P^-_y(P^-_\delta)$. are realized by the action of the supermirror (analyzer), while applying the respective magnetic fields in DC-coil 2. (b) Bloch sphere description of projectors $P^+_z,P^+_y,P^+_\delta$ and $P_z^-,P^-_y,P^-_\delta$.} 
	\label{fig:SetUp}
\end{SCfigure*}


\section{Experimental Setup and Procedure}%
In this letter, we present a neutron optical test of the tight state-independent preparations uncertainties described by Eqs.\,(\ref{eq:ExpVal}), (\ref{eq:StDec}) and (\ref{eq:Entropy}). The experiment was carried out at the polarimeter instrument \emph{NepTUn (NEutron Polarimeter TU wieN)}, located at the tangential beam port of the 250\,kW TRIGA research reactor at the Atominstitut - TU Wien, in Vienna, Austria. A schematic illustration of the experimental setup is depicted in Fig.\,\ref{fig:SetUp}. An incoming monochromatic neutron beam with mean wavelength $\lambda\simeq 2.02\,\AA$ ($\Delta\lambda/\lambda\simeq0.02$) is polarized along the vertical  ($+z$) direction by refraction from a tunable CoTi multilayer array, hence on referred to as supermirror. The incident (mixed) state is given by $\rho_z^{\rm{in}} =\frac{1}{2}({1\!\!1}+r\sigma_z)$, where $r=1$ corresponds to the pure state $\vert+z\rangle$. The mixing parameter $r$ is is adjusted by the incident angle $\beta_z^r$ between the supermirror and the neutron beam, in the required parameter region. Experimentally, initial degrees of polarizations between $r_{\rm{min}}=0.83(1)$ and $r_{\rm{max}}=0.99(1)$ were achieved. To avoid depolarization, the setup is covered by a 13\,G guide field in vertical direction (not depicted in Fig.\,\ref{fig:SetUp}). The initial states are chosen by a classical randomness generator (RG) and prepared by direct current (DC) coil 1, which generates a static magnetic field $B_x^\theta$ pointing to the $x$-direction. The magnetic field induces a unitary Larmor precession $U_{\rm{DC}}=e^{i\theta\sigma_x}$ by an angle $\theta$ inside the coil expressed as 
\begin{eqnarray}
\rho_{\rm{in}}^\theta=U^\dagger_{\rm{DC}}\rho_z^{\rm{in}} U_{\rm{DC}}. 
\end{eqnarray}
The angle of rotation $\theta = \gamma B_x t$ is proportional to the magnetic field strength and the time $t$ of passage of the neutron through the coil, $\gamma$ being the gyromagnetic ratio $\gamma=-\frac{2\vert\mu\vert}{\hbar}$, where $\mu$ denotes the magnetic moment of the neutron. Since the transition time $t$ is constant, the polar angle $\theta$ of the initial spin state $\rho_{\rm{in}}^\theta$ is entirely controlled by the electric current in the coil that generates the magnetic field~$B_x$. The azimuthal angle $\varphi$ of the prepared initial state $\rho_{\rm{in}}^{\theta,\varphi}$, given by  
\begin{eqnarray}
\rho_{\rm{in}}^{\theta,\varphi}=U^\dagger_{\rm{GF}}U^\dagger_{\rm{DC}}\rho_z^{\rm{in}} U_{\rm{DC}}U_{\rm{GF}}
\end{eqnarray}
 and $U_{\rm{GF}}=e^{i\varphi\sigma_z}$, is induced by Larmor precession within the static magnetic guide field (GF). The respective angle is adjusted by the appropriate position $y_0^\varphi$ of DC-coil 1. Note that all randomly selected initial states lie on the boundary region of the respective tight uncertainty relation and belong to a subset of all possible states.

Our experimental test of the tight uncertainty relations Eqs.\,(\ref{eq:ExpVal}-\ref{eq:Entropy}) is conducted for two fixed Pauli operators $A=\vec a\cdot\vec\sigma$ and $B=\vec b\cdot\vec \sigma$, with (i) $\vec a\cdot \vec b=0$ and (ii) $\vec a\cdot \vec b=\frac{1}{2}$. For (i) we chose $\vec a=(0,0,1)^T$ and $\vec b=(0,1,0)^T$, thus the four projectors $P^+_z,P^-_z,P^+_y$ and $P^-_y$ were measured for every randomly chosen initial state $\rho_{\rm{in}}^{\theta,\varphi}$, resulting in an observed intensity denoted as $I={\rm{Tr}}(\rho_{\rm{in}}^{\theta,\varphi}\,P_i^j)$, with $i=z,y$ and $j=+,-$. The projectors are realized by the action of the supermirror (analyzer) while applying the respective magnetic fields in DC-coil 2. Inside DC-coil 2, the magnetic field $B_{x}^\theta$ induces spinor rotations of $\vartheta=0,\pi,+\frac{\pi}{2}$ and $\vartheta=-\frac{\pi}{2}$ about the $x$-axis, required for projective measurements along the $+z,-z,+y$ and $-y$-direction, respectively. Since all four projectors lie in the $y$-$z$-plane the position of DC-coil 2 remains unchanged. For (ii) $\vec a\cdot \vec b=\frac{1}{2}$, which corresponds to an angle of $\vartheta=60\,$deg between $\vec a$ and $\vec b$, we have again $\vec a=(0,0,1)^T$ but $\vec b=(0,\frac{\sqrt 3}{2},\frac{1}{2})^T$ now. The respective projectors are denoted as $P^+_z, P^-_z,$ $P^+_\delta$ and $P^-_\delta$, respectively.
\section{Experimental Results}

\begin{figure}[!b]
	\includegraphics[width=0.23\textwidth]{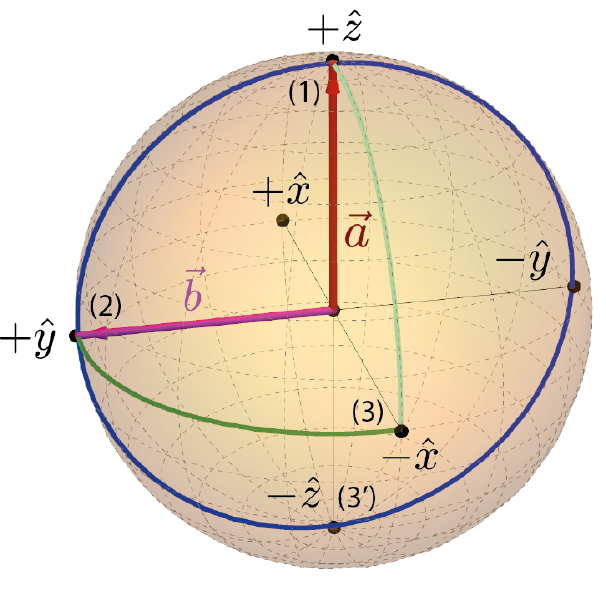}
	\caption{Bloch sphere of initial states saturating boundaries of Eqs.\,(\ref{eq:ExpVal}), (\ref{eq:StDec}) and (\ref{eq:Entropy}), for expectations values, standard deviations and entropies, respectively, in case $\vec a\cdot \vec b=0$. } 
	\label{fig:Bloch1}
\end{figure}

\begin{figure*}
	\includegraphics[width=0.97\textwidth]{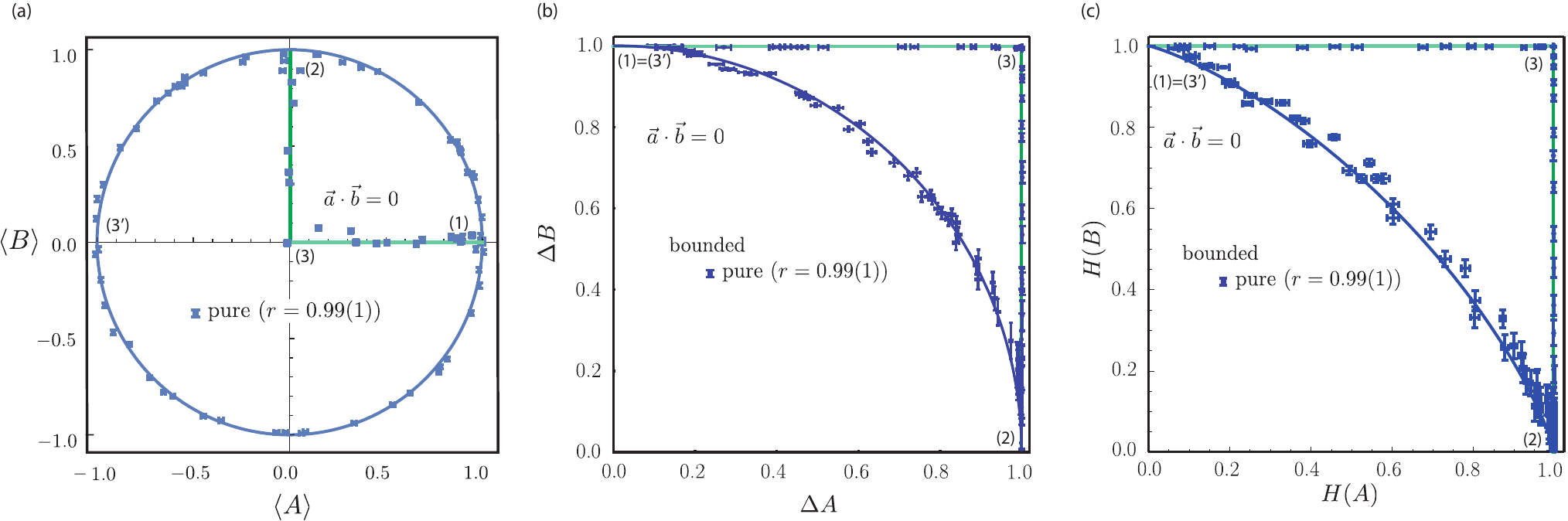}
	\caption{Plots of the experimentally obtained values for Pauli observables $\vec a\cdot \vec b=0$, in terms of expectation values $\langle A \rangle$, $\langle B \rangle$ in (a), standard deviations $\Delta A$, $\Delta B$ in (b), and entropies $H(A)$, $H(B)$ in (c). Blue curves indicate the theoretic predictions of lower bounds from Eqs.(\ref{eq:ExpVal}),\,(\ref{eq:StDec}) and (\ref{eq:Entropy}), for expectation values, standard deviations and entropies, respectively. Dark and light green lines represent the theoretic predictions for values of the corresponding initial states indicated by the respective color on the Bloch sphere in Fig.\,\ref{fig:Bloch1}.} 
	\label{fig:ExStH}
\end{figure*}


\subsection{State-independent relations}\label{sec:state}


The lowest or \emph{state-independent} bound, expressed by the very RHS of Eqs. (\ref{eq:ExpVal}), (\ref{eq:StDec}) and (\ref{eq:Entropy}), for expectation values, standard deviations and entropies, respectively, is saturated for \emph{pure} initial states ($r=1$), which will be studied first.

\vspace{-7mm}

\subsubsection{Configuration $\vec a\cdot \vec b=0$ }\label{sec:PureEV}
Two Pauli observables $A=\vec a\cdot\vec\sigma$ and $B=\vec b\cdot\vec\sigma$, with $\vec a=+\hat z$ and $\vec b=+\hat y$, yielding  $\vec a\cdot \vec b=0$, are selected (see Fig.\,\ref{fig:Bloch1}). Pure initial states, located on the great circle in the plane spanned by the observables' unit vectors $\vec a$ and $\vec b$, depicted in blue on the Bloch sphere in Fig.\,\ref{fig:Bloch1}, form the lower bound of allowed values. However, the lower bound is not given by a closed curve in case of standard deviations and entropies. Therefore, additional initial states, indicated by the green and light green arcs on the Bloch sphere, are required to close the boundary of all allowed values. \\
i) Expectation values (EV): equation (\ref{eq:ExpVal}) sets tight constrains on the allowed values for the expectation values $\langle A \rangle$ and $\langle B \rangle$ which is experimentally tested with a set of randomly chosen initial states $\rho_{\rm{in}}^\theta$. The state-independent bound of Eq.\,(\ref{eq:ExpVal}), given by $b_{\rm{EV}}=\vert \vec a \times\vec b\vert^2$, is saturated only by pure states, distributed on the \emph{great circle} connecting north and south pole of the Bloch sphere via $+y$. This great circle, depicted in blue in Fig.\,\ref{fig:Bloch1}, is embedded in the plane spanned by the observables' unit vectors $\vec a$ and $\vec b$ and parameterized by the polar angle $\theta \in [0,2\pi]$ and $\varphi=\frac{\pi}{2}$. There is a one-to-one correspondence between the initial states' polar angle $\theta$ and the angle on the circle forming the boundary of allowed values for expectation values of $\langle A\rangle$ and $\langle B\rangle$, plotted in Fig.\,\ref{fig:ExStH}\,(a). Therefore, in the actual experiment the position of DC-coil 1 remains fixed for this measurement. Starting at the north pole ($\theta=0$), indicated as point (1) in Fig.\,\ref{fig:Bloch1}, we have $\langle A\rangle=1$ and $\langle B\rangle=0$. At $\theta=\frac{\pi}{2}$ ($+y$-direction), indicated by point (2), the situation reverses with $\langle A\rangle=0$ and $\langle B\rangle=1$. Closing the great circle on the Bloch sphere from $\theta=\frac{\pi}{2}$ to $\theta=2\pi$ yields a closed curve for the boundary of all possible values of expectation values $\langle A\rangle$ and $\langle B\rangle$. Initial states outside the blue great circle, for instance states connecting points (2) and (3) - light green states in Fig.\,\ref{fig:Bloch1} -  are unbounded pure states, as seen from Fig.\,\ref{fig:ExStH}\,(a). At point (3) ($-x$-direction) expectation values yield $\langle A\rangle=\langle B\rangle=0$, and are therefore found at the origin, which is the center of the region of allowed values for expectation values $\langle A\rangle$ and $\langle B\rangle$.
\begin{figure*}
	\includegraphics[width=0.99\textwidth]{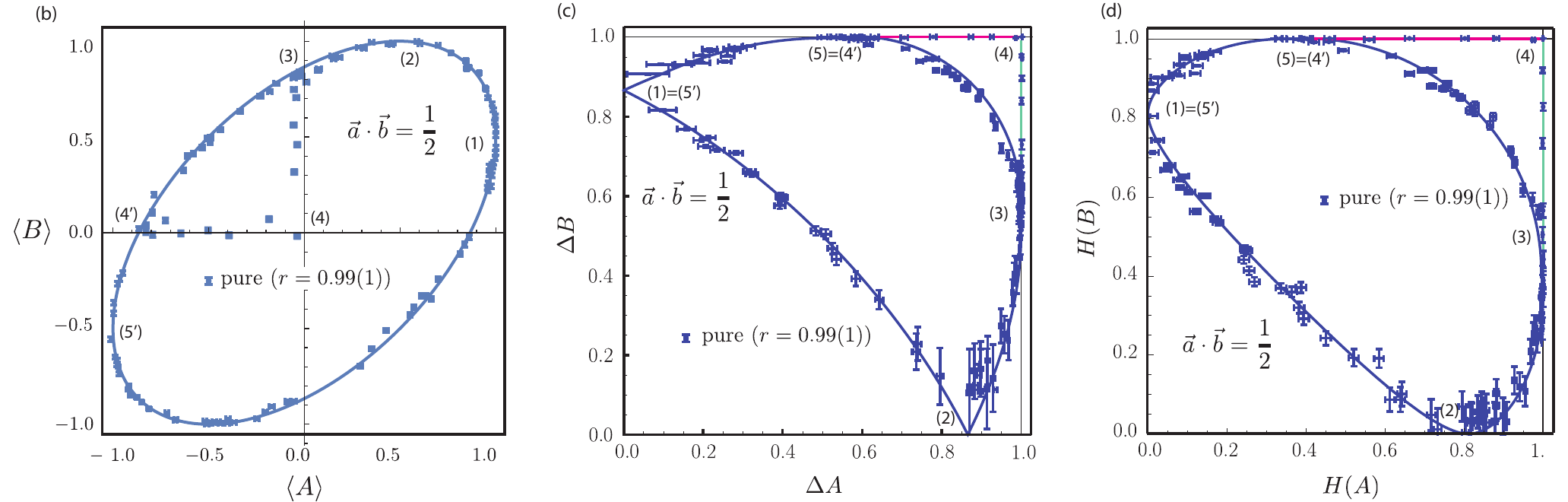}
	\caption{lots of the experimentally obtained values for Pauli observables $\vec a\cdot \vec b=\frac{1}{2}$, in terms of expectation values $\langle A \rangle$, $\langle B \rangle$ in (a), standard deviations $\Delta A$, $\Delta B$ in (b), and entropies $H(A)$, $H(B)$ in (c). Blue curves indicate the theoretic predictions of lower bounds from Eqs.(\ref{eq:ExpVal}),\,(\ref{eq:StDec}) and (\ref{eq:Entropy}), for expectation values, standard deviations and entropies, respectively. Light green and magenta  lines represent the theoretic predictions for values of the corresponding initial states indicated by the respective color on the Bloch sphere in Fig.\,5.}
	\label{fig:UnTight2}
\end{figure*}

ii) Standard deviations (SD): since expectation value and standard deviation of Pauli observables are one-to-one related via $\Delta X=\sqrt{\langle X\rangle^2-1}$, the data obtained from above is accordingly transformed to evaluate the tight state-independent preparation uncertainty relations as expressed in Eq.\,(\ref{eq:StDec}), with lower state-independent bound $b_{\rm{SD}}=1+(\vec a\cdot\vec b)^2$. Unlike in the case of expectation values, pure states on the great circle in the $y$-$z$ plane saturate only the (state-independent) lower bound (curved boundary) but do not cover the entire region of allowed values for standard deviations $\Delta A$ and  $\Delta B$, which can be seen in  Fig.\,\ref{fig:ExStH}\,(b). At $\theta=0$ (point (1), $+z$-direction), standard deviation $\Delta B$ starts at maximal value (for $r=1$ this is $\Delta B=1$) and  $\Delta A$ is minimal (for $r=1$ this is $\Delta A=0$). For increasing values of $\theta$ (while keeping $\varphi=\frac{\pi}{2}$ constant) $\Delta B$ decreases while $\Delta A$ increases. At $\theta=\frac{\pi}{2}$ (point (2), $+y$-direction) $\Delta B$ is minimal (for $r=1$, $\Delta B=0$) and $\Delta A$ is maximal (for $r=1$, $\Delta A=1$). In the interval  $\theta\,\in\,[\frac{\pi}{2},\pi]$ the reverse behavior is observed and at $(\theta,\varphi)=(\frac{\pi}{2},\pi)$ (point (3), $-x$-direction), we have again $\Delta A=0$ and $\Delta B=1$ (as for $\theta=0$). For $\theta\,\in\,[\pi,2\pi]$, the results of $\theta\,\in\,[0,\pi]$ are reproduced. The vertical boundary, corresponding to a constant (maximal) value of $\Delta A=1$, is obtained for initial states $\rho_{\rm{in}}^{\theta,\varphi}$ with constant polar angle $\theta=\frac{\pi}{2}$ and randomly generated azimuthal angle $\varphi\,\in\,[\frac{\pi}{2},\pi]$, these states are found on the dark green region of the equatorial plane of the Bloch sphere in Fig.\,\ref{fig:Bloch1}. For $(\theta,\varphi)=(\frac{\pi}{2},\pi)$, point (3) the upper right corner with $\Delta A=\Delta B=1$ is reached. For the horizontal boundary ($\Delta B=1$), $\varphi$ is kept constant at $\varphi=\pi$, while $\theta$ is randomly chosen from the interval $[0,\frac{\pi}{2}]$ (light green curve on the Bloch sphere), where for $\theta=0$ ($+z$-direction) the boundary becomes a closed curve. 

\begin{figure}[!b]
	\includegraphics[width=0.26\textwidth]{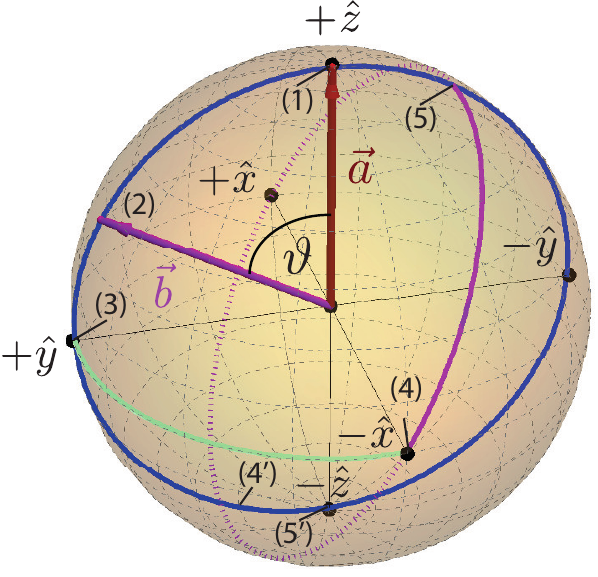}
	\caption{Bloch sphere of initial states saturating boundaries of Eqs.\,(\ref{eq:ExpVal}), (\ref{eq:StDec}) and (\ref{eq:Entropy}), for expectations values, standard deviations and entropies, respectively, in case $\vec a\cdot \vec b=\frac{1}{2}$.} 
	\label{Bloch2}
\end{figure}

iii) Entropies: the approach presented in \cite{Abbott16} for a tight state-independent uncertainty relations for qubits, is based on the fact that in the case of Pauli observables the expectation value contains all information necessary to derive the uncertainty. Consequently, the uncertainty can also be expressed  in terms of entropy $H$, which is also a function of the expectation value. The lower (state-independent) bound is calculated as $b_{ H}=1-(\vec a\cdot\vec b)^2$. All arguments on the initial states saturating the boundary for standard deviations $\Delta A$ and $\Delta B$ also apply to entropies $H(A)$ and $H(B)$, plotted in  Fig.\,\ref{fig:ExStH}\,(c).


\subsubsection{Configuration $\vec a\cdot \vec b=\frac{1}{2}$ }

Next, expectation values, standard deviations and entropies (depicted in Fig.\,\ref{fig:UnTight2}) for Pauli observables $A=\vec a\cdot\vec\sigma$ and $B=\vec b\cdot\vec\sigma$, with $\vec a\cdot \vec b=\frac{1}{2}$, which corresponds to a relative angle $\vartheta=60$\,deg (see Fig.\,\ref{Bloch2}), are investigated.

i) Expectation value: the obtained values for $\langle A \rangle$ and $\langle B \rangle$ now have an elliptical boundary, which is depicted in Fig.\,\ref{fig:UnTight2} (a). For pure states with $\theta=0$ (point (1), $+z$-direction), neither of the two expectation vales $\langle A \rangle$ and $\langle B \rangle$ is zero, more precisely $\langle A \rangle=1$ and $\langle B \rangle=\frac{1}{2}$. For increasing values of $\theta$ (while keeping $\varphi=\frac{\pi}{2}$ constant) $\langle A \rangle=1$ decreases, while $\langle B \rangle=1$ increases, reaching a maximum of $\langle B \rangle=1$ (with $\langle A\rangle=\frac{1}{2}$) at $\theta=\frac{\pi}{3}=\vartheta$, that is the polar angle of unit vector $\vec b$, indicated by point (2) on the Bloch sphere in Fig.\,\ref{Bloch2}. In the interval $\theta\in[\frac{\pi}{3},\frac{\pi}{2}]$ both  $\langle A \rangle$ and $\langle B \rangle$  are decreasing. At $\theta=\frac{\pi}{2}$ (point (3), $+y$-direction) $\langle A \rangle=0$ and $\langle B \rangle= 0.87$. Polar angle $\theta=\frac{5\pi}{6}$ yields $\langle B \rangle=0$ and $\langle A \rangle= -0.87$,  at point (4'). A minimum for $\langle A \rangle$ is reached at $\theta=\pi$ (point (5'), $-z$-direction) with $\langle A \rangle=-1$ (and $\langle B \rangle=-\frac{1}{2}$ ). In the interval  $\theta\,\in\,[\pi,2\pi]$ the reverse behavior is observed.  Initial states outside the blue great circle, for instance states connecting points (3) and (4), are unbounded pure states.

\vspace{2mm}

ii) Standard deviations: initial states that saturate the state-independent lower bound of Eq.\,(\ref{eq:StDec}) (curved boundaries from point (5) to point (3) in Fig.\,\ref{fig:UnTight2} (b)) are located on the blue great circle in the $z$-$y$-plane of Fig.\,\ref{fig:UnTight2} (a) with polar angle $\theta\,\in\,[-\frac{\pi}{6},\frac{\pi}{2}]$. For pure states with $\theta=0$ (point (1), $+z$-direction) $\Delta A=0$ ($\Delta B= 0.87$) is obtained. At $\theta=\frac{\pi}{3}=\vartheta$ point (2) we have $\Delta B=0$ ($\Delta A= 0.87$) and  at $\theta=\frac{\pi}{2}$, point (3), $\Delta A=1$ and $\Delta B=\frac{1}{2}$. The vertical boundary, represented by points (3) to (4) is covered by initial states on the equatorial plane of the Bloch sphere with azimuthal angle $\varphi \in [\frac{\pi}{2},\pi]$ (light green line in Fig.\,\ref{fig:UnTight2} (b)). Initial states saturating the horizontal lower bound (magenta in Fig.\,\ref{fig:UnTight2} (b)) are located on a great circle (magenta in Fig.\,\ref{Bloch2}) embedded in a plane perpendicular to $\vec{b}$. Here both polar angle $\theta$ and azimuthal angle $\varphi$ are varied, namely $\theta$ between $\frac{\pi}{2}$ and $-\frac{\pi}{6}$ and $\varphi$ between $\pi$ and $\frac{\pi}{2}$, before reaching point (4), thereby closing the boundary of allowed values for standard deviations $\Delta A$ and $\Delta B$. Initial states on the blue great circle in the $z$-$y$ plane of Fig.\,\ref{Bloch2} ($\varphi=\frac{\pi}{2}$) with $\theta>\frac{\pi}{2}$ are located inside the boundary (unbounded pure states), more precisely on the blue curve in Fig.\,\ref{fig:UnTight2} (b) between point (3) and point (4'). At $\theta=\frac{5\pi}{6}$, point (4') is reached, yielding the same values as $\theta=-\frac{\pi}{6}$ of point (5).

iii) Entropies: initial states saturating the boundary for entropies $H(A)$ and $H(B)$ are again the same as for standard deviations $\Delta A$ and $\Delta B$, which is plotted in Fig.\,\ref{fig:UnTight2}\,(c). Values for  entropies $H(A)$ and $H(B)$ in point ($i$) denoted as $\{ {\rm p}(i);H(A),H(B)\}$ are given by $\{ {\rm p}(1);0,\,0.81\}$, $\{ {\rm p}(2);0.81,0\}$, $\{ {\rm p}(3);1,\,0.35\}$, $\{ {\rm p}(4);1,1\}$, and $\{ {\rm p}(5);0.35,1\}$.

\subsection{Partially state-dependent relations}\label{sec:statedep}
\subsubsection{Expectation Values}

\begin{figure}[!t]
	\includegraphics[width=0.4\textwidth]{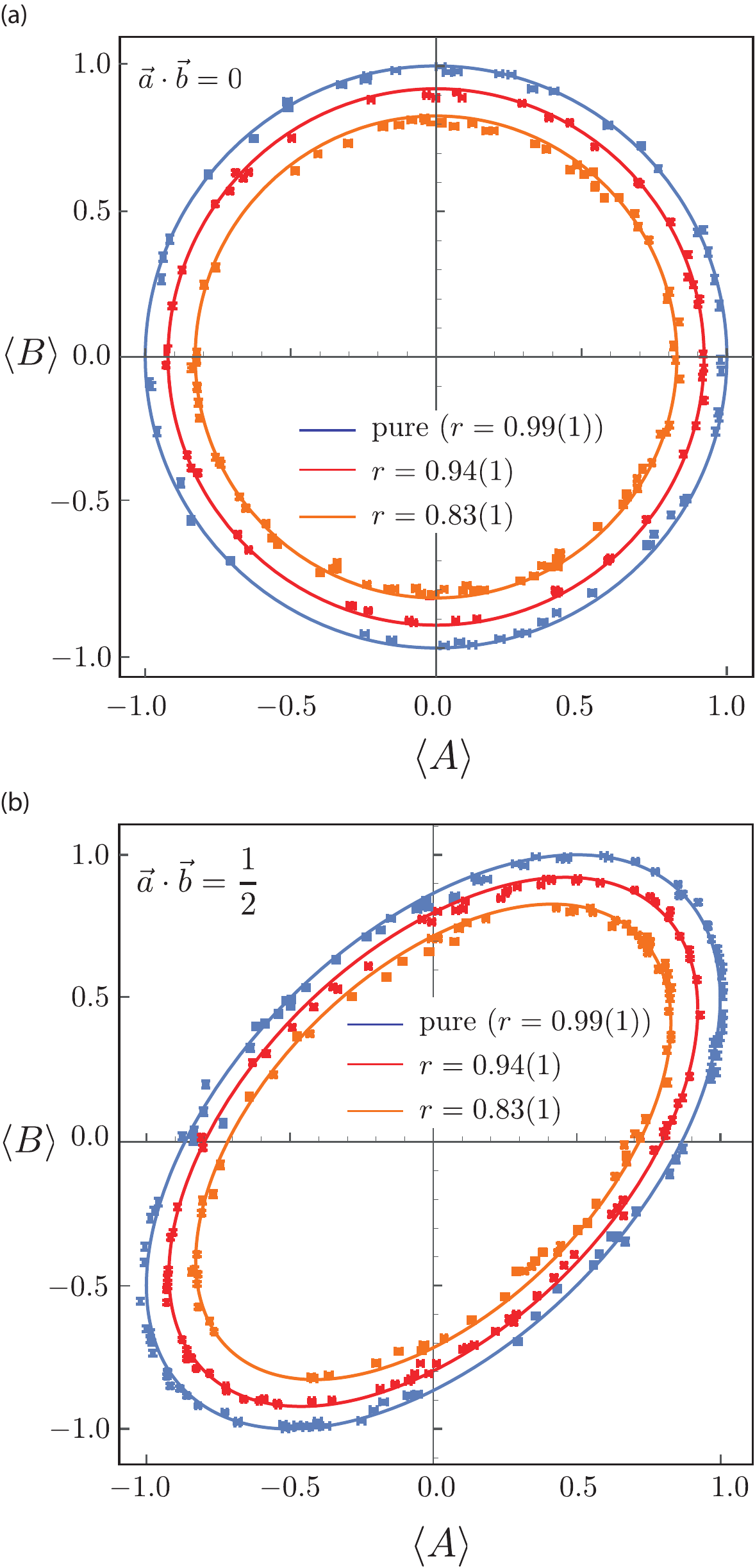}
	\caption{Plots of the experimentally obtained values for expectation values $\langle A \rangle$ and $\langle B \rangle$ for (a) $\vec a\cdot\vec b=0$ and (b) $\vec a\cdot\vec b=\frac{1}{2}$. The colored ellipses indicate the lower bounds of Eq.(\ref{eq:ExpVal}), of allowed values for $\langle A \rangle$ and $\langle B \rangle$ for three different initial degrees of polarization. }
	\label{fig:PlotResultsEx}
\end{figure}
\vspace{2mm}

As already discussed in Sec.\,\ref{sec:PureEV}, the state-independent bound of Eq.\,(\ref{eq:ExpVal}), given by $b_{\rm{EV}}=\vert \vec a \times\vec b\vert^2$, is saturated \emph{only} by \emph{pure} states, found on the surface of the $y$-$z$-plane on the Bloch sphere. The \emph{partially} state-dependent lower bound, expressed as $b'_{\rm{EV}}(r)=(1-(\vec a\cdot\vec b)^2)\vert \vec r\vert^2$, is covered by \emph{mixed states} located in the $y$-$z$-plane of the Bloch sphere, with respective degree of polarization $r$. For expectation values the lower bound of Eq.\,(\ref{eq:ExpVal}) is a closed curve representing the entire boundary of allowed values for $\langle A \rangle$ and $\langle B \rangle$, which can be seen in Fig.\,\ref{fig:PlotResultsEx} (a) and \ref{fig:PlotResultsEx} (b), for $\vec a\cdot\vec b=0$ and $\vec a\cdot\vec b=\frac{1}{2}$, respectively. The measurement is carried out for three initial degrees of polarization, which are tuned by the angle $\beta_z^r$ between the supermirror and the neutron beam, namely $r_{\rm{min}}=0.83(1)$, $r_{\rm{mid}}=0.94(1)$ and $r_{\rm{max}}=0.99(1)$. For all initial degrees of polarization the theoretical predictions for expectations vales $\langle A \rangle$ and $\langle B \rangle$ (solid lines in Fig.\,\ref{fig:PlotResultsEx}) are reproduced evidently.

\subsubsection{Standard Deviations}%
\begin{figure}[!t]
	\includegraphics[width=0.4\textwidth]{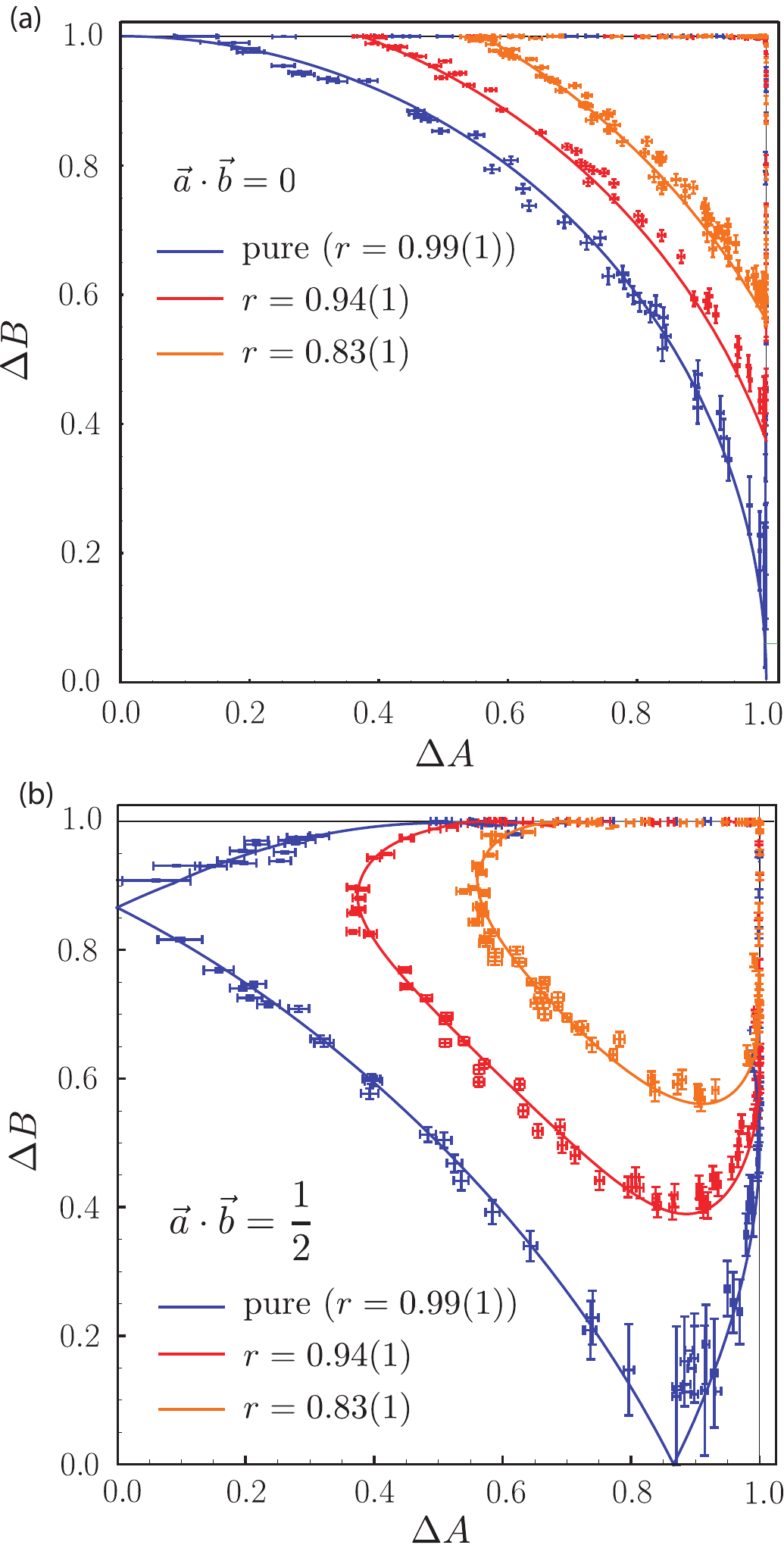}
	\caption{Plots of the experimentally obtained values for standard deviations $\Delta A$ and $\Delta B$ for (a) $\vec a\cdot\vec b=0$ and (b) $\vec a\cdot\vec b=\frac{1}{2}$. The colored curves indicate the lower bounds of Eq.(\ref{eq:StDec}) of allowed values for $\Delta A$ and $\Delta B$ for three different initial degrees of polarization.}
	\label{fig:PlotResultsStDev}
\end{figure}
All states that saturate the state-independent bound of Eq.\,(\ref{eq:StDec}), denoted as $b_{\rm{SD}}=1+(\vec a\cdot\vec b)^2$, are \emph{pure} states, found on the surface of the $y$-$z$-plane on the Bloch sphere. For $\vec a\cdot\vec b=0$ the partially state-dependent lower bound of Eq.\,(\ref{eq:StDec}), that is $b'_{\rm{SD}}(r)=2-\big(1-(\vec a\cdot\vec b)^2\big)\vert\vec r\vert^2$, is saturated by the corresponding mixed states in the $y$-$z$ plane of the Bloch sphere with polar angle $\theta\in [0,\pi/2]$. Unlike the case of expectation values, the lower bound of Eq.\,(\ref{eq:StDec}) is not a closed curve, which can be seen Fig.\,\ref{fig:PlotResultsStDev}. While initial states $\rho_{\rm{in}}^\theta$, that lie in the $y$-$z$-plane, cover the entire the lower bound of Eq.\,(\ref{eq:ExpVal}), they are insufficient to enclose the remaining boundaries (vertical and horizontal lines in Fig.\,\ref{fig:PlotResultsStDev}) of allowed values for standard deviations $\Delta A$ and $\Delta B$. For standard deviations, the situation is different compared to expectation values; the vertical and horizontal boundaries can not only be saturated by pure states (which cover again the entire bound), but (partially) also by certain mixed states ($r<1$). The vertical and horizontal boundary of allowed values is occupied by initial states of all mixing angles. For $\vec a\cdot \vec b=\frac{1}{2}$, depicted in Fig.\,\ref{fig:PlotResultsStDev} (b), the partially state-dependent lower bounds of Eq.\,(\ref{eq:StDec}) (curved boundaries), are obtained for pure and mixed  initial states $\rho_{\rm{in}}^\theta$ with $\theta\,\in\,[-\frac{\pi}{6},\frac{\pi}{2}]$, which are randomly generated. For all three initial degrees of polarization ($r_{\rm{min}}=0.83$, $r_{\rm{mid}}=0.94$ and $r_{\rm{max}}=0.99$) the theoretical predictions of the tight state-independent and tight partially state-dependent uncertainty relations in terms of standard deviations $\Delta A$ and $\Delta B$ (solid lines in Fig.\,\ref{fig:PlotResultsStDev}) are experimentally confirmed. 
 \vspace{3mm}
\begin{figure}[!t]
	\includegraphics[width=0.4\textwidth]{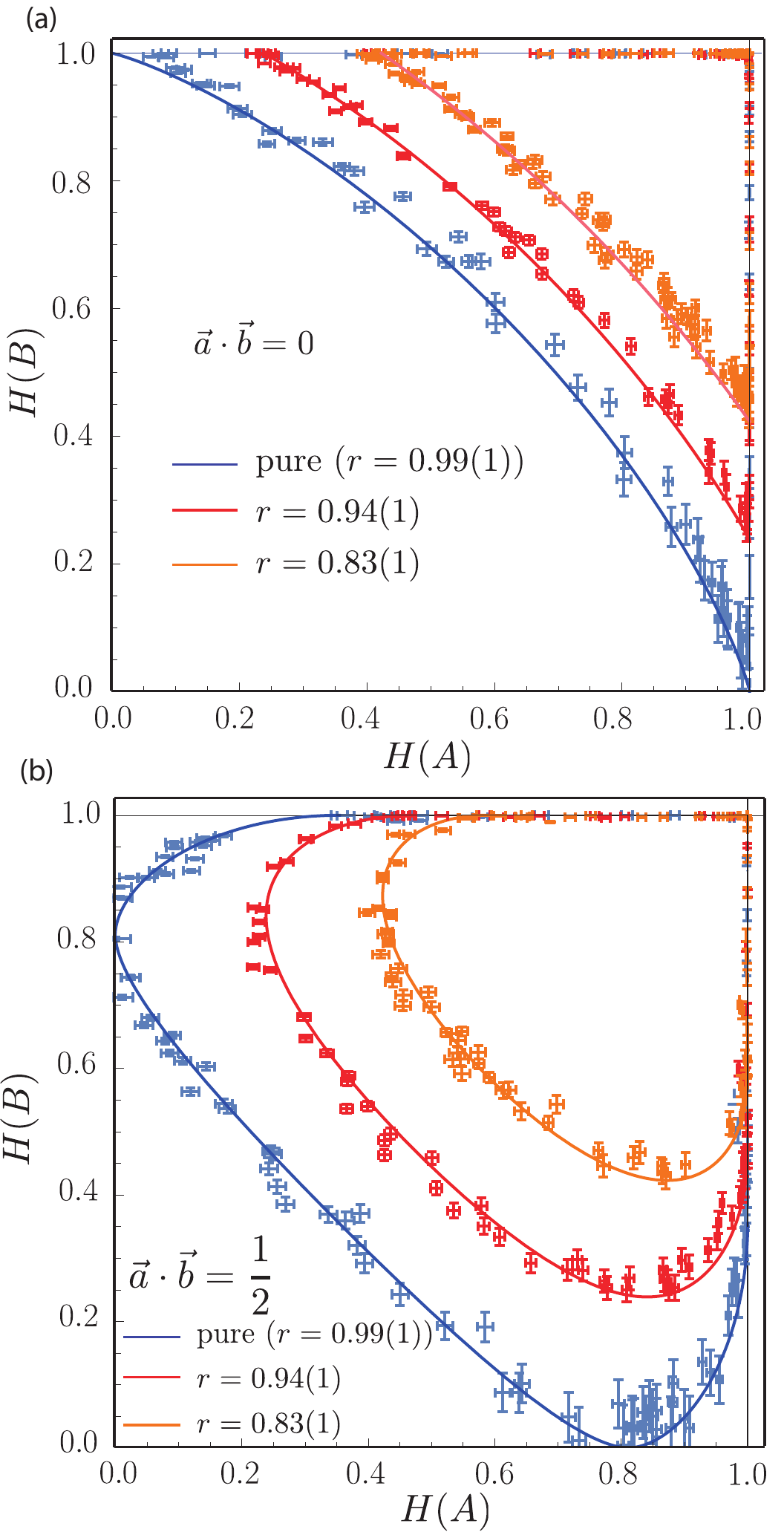}
	\caption{Plots of the experimentally obtained values for entropies $H(A)$ and $H(B)$ for (a) $\vec a\cdot\vec b=0$ and (b) $\vec a\cdot\vec b=\frac{1}{2}$. The colored curves indicate the lower bounds of Eq.(\ref{eq:Entropy}) of allowed values for $H(A)$ and $H(B)$ for three different initial degrees of polarization.}
	\label{fig:PlotResultsEntropy}
\end{figure}

\subsubsection{Entropy}%
The obtained results for three initial degrees of polarizations ($r_{\rm{min}}=0.83$, $r_{\rm{mid}}=0.94$ and $r_{\rm{max}}=0.99$) are depicted in Fig.\,\ref{fig:PlotResultsEntropy}. Again, as in the case of standard deviations for $\vec a\cdot \vec b=0$, depicted in Fig.\,\ref{fig:PlotResultsStDev} (a), all states that saturate the state-independent bound of Eq.\,(\ref{eq:Entropy}), denoted as $b_{{H}}=1-(\vec a\cdot\vec b)^2$, are \emph{pure} states, located on the surface of the $y$-$z$ plane on the Bloch sphere with polar angle $\theta\in [0,\frac{\pi}{2}]$. The partially state-dependent lower bound of Eq.\,(\ref{eq:Entropy}), expressed as $b'_{{H}}(r)=\big(1-(\vec a\cdot\vec b)^2\big)\vert\vec r\vert^2$, is saturated by the corresponding mixed states in the $y$-$z$ plane of the Bloch sphere with polar angle  $\theta\in [0,\frac{\pi}{2}]$. For $\vec a\cdot \vec b=\frac{1}{2}$, depicted in Fig.\,\ref{fig:PlotResultsEntropy} (b), the bounds of Eq.\,(\ref{eq:Entropy}), are obtained for pure and mixed initial states $\rho_{\rm{in}}^\theta$ with $\theta\,\in\,[-\frac{\pi}{6},\frac{\pi}{3}]$. The theoretical predictions, indicated by solid lines in Fig.\,\ref{fig:PlotResultsEntropy} are reproduced evidently, demonstrating tight state-independent and tight partially state-dependent uncertainty relations for entropies $H (A)$ and $H(B)$. 

\section{Discussion and Conclusion}
The presented experiment investigates the relationship between the expectation values of Pauli spin observables and two standard
measures of uncertainty, namely standard deviations and Shannon entropies. The tightness of state-independent  uncertainty relations for Pauli measurements on qubits is experimentally demonstrated. In addition, we observed bounds on these relations, expressed in terms of the norm $|r|$ of the Bloch vector, resulting in (partially) state-dependent uncertainty relations with lower bounds. We have experimentally confirmed the tightness of state-independent, as well as partially state-dependent, uncertainty relations for pairs of Pauli measurements on qubits. The observed uncertainty relations, expressed in terms of standard deviations and Shannon entropy (both functions of the expectation value), completely characterize the allowed values of uncertainties for Pauli spin observables. The theoretical framework allows for uncertainty relations for three (or more) observables, which will be a topic of forthcoming publications. Finally, we want to emphasize, that it is also possible to go beyond projective measurements and give similar relations for positive-operator valued measures (POVMs)
for qubits with binary outcomes, which will be investigated in upcoming experiments.

\begin{acknowledgments}
The authors thank Alastair A. Abbott and Cyril Branciard for helpful discussions. S.S.\ and Y.H.\ acknowledge support by the Austrian science fund (FWF) Projects No.\ P30677-N20 and No.\ P27666-N20.
 \end{acknowledgments}


\begin{thebibliography}{30}
\expandafter\ifx\csname natexlab\endcsname\relax\def\natexlab#1{#1}\fi
\expandafter\ifx\csname bibnamefont\endcsname\relax
  \def\bibnamefont#1{#1}\fi
\expandafter\ifx\csname bibfnamefont\endcsname\relax
  \def\bibfnamefont#1{#1}\fi
\expandafter\ifx\csname citenamefont\endcsname\relax
  \def\citenamefont#1{#1}\fi
\expandafter\ifx\csname url\endcsname\relax
  \def\url#1{\texttt{#1}}\fi
\expandafter\ifx\csname urlprefix\endcsname\relax\def\urlprefix{URL }\fi
\providecommand{\bibinfo}[2]{#2}
\providecommand{\eprint}[2][]{\url{#2}}

\bibitem[{\citenamefont{Kennard}(1927)}]{Kennard27}
\bibinfo{author}{\bibfnamefont{E.~H.} \bibnamefont{Kennard}},
  \bibinfo{journal}{Z. Phys.} \textbf{\bibinfo{volume}{44}},
  \bibinfo{pages}{326} (\bibinfo{year}{1927}).

\bibitem[{\citenamefont{Heisenberg}(1927)}]{Heisenberg27}
\bibinfo{author}{\bibfnamefont{W.}~\bibnamefont{Heisenberg}},
  \bibinfo{journal}{Z. Phys.} \textbf{\bibinfo{volume}{43}},
  \bibinfo{pages}{172} (\bibinfo{year}{1927}).

\bibitem[{\citenamefont{Robertson}(1929)}]{Robertson29}
\bibinfo{author}{\bibfnamefont{H.~P.} \bibnamefont{Robertson}},
  \bibinfo{journal}{Phys. Rev.} \textbf{\bibinfo{volume}{34}},
  \bibinfo{pages}{163} (\bibinfo{year}{1929}).

\bibitem[{\citenamefont{Schr{\"o}dinger}(1930)}]{Schroedinger30}
\bibinfo{author}{\bibfnamefont{E.}~\bibnamefont{Schr{\"o}dinger}},
  \bibinfo{journal}{Sitzungsberichte der Preussischen Akademie der
  Wissenschaften, Physikalisch-mathematische Klasse}
  \textbf{\bibinfo{volume}{14}}, \bibinfo{pages}{296} (\bibinfo{year}{1930}),
  \bibinfo{note}{engl. translation at http://arxiv.org/abs/quant-ph/9903100}.

\bibitem[{\citenamefont{Ozawa}(2003{\natexlab{a}})}]{Ozawa03}
\bibinfo{author}{\bibfnamefont{M.}~\bibnamefont{Ozawa}},
  \bibinfo{journal}{Phys. Rev. A} \textbf{\bibinfo{volume}{67}},
  \bibinfo{pages}{042105} (\bibinfo{year}{2003}{\natexlab{a}}).

\bibitem[{\citenamefont{Busch et~al.}(2013)\citenamefont{Busch, Lahti, and
  Werner}}]{Busch13}
\bibinfo{author}{\bibfnamefont{P.}~\bibnamefont{Busch}},
  \bibinfo{author}{\bibfnamefont{P.}~\bibnamefont{Lahti}}, \bibnamefont{and}
  \bibinfo{author}{\bibfnamefont{R.~F.} \bibnamefont{Werner}},
  \bibinfo{journal}{Phys. Rev. Lett.} \textbf{\bibinfo{volume}{111}},
  \bibinfo{pages}{160405} (\bibinfo{year}{2013}).

\bibitem[{\citenamefont{Branciard}(2013)}]{Branciard13}
\bibinfo{author}{\bibfnamefont{C.}~\bibnamefont{Branciard}},
  \bibinfo{journal}{Proc. Natl. Acad. Sci. USA} \textbf{\bibinfo{volume}{17}},
  \bibinfo{pages}{6742} (\bibinfo{year}{2013}).

\bibitem[{\citenamefont{Ozawa}(2014)}]{Ozawa14}
\bibinfo{author}{\bibfnamefont{M.}~\bibnamefont{Ozawa}},
  \bibinfo{journal}{arXiv:1404.3388v1 [quant-ph]}  (\bibinfo{year}{2014}).

\bibitem[{\citenamefont{Erhart et~al.}(2012)\citenamefont{Erhart, Sponar,
  Sulyok, Badurek, Ozawa, and Hasegawa}}]{Erhart12}
\bibinfo{author}{\bibfnamefont{J.}~\bibnamefont{Erhart}},
  \bibinfo{author}{\bibfnamefont{S.}~\bibnamefont{Sponar}},
  \bibinfo{author}{\bibfnamefont{G.}~\bibnamefont{Sulyok}},
  \bibinfo{author}{\bibfnamefont{G.}~\bibnamefont{Badurek}},
  \bibinfo{author}{\bibfnamefont{M.}~\bibnamefont{Ozawa}}, \bibnamefont{and}
  \bibinfo{author}{\bibfnamefont{Y.}~\bibnamefont{Hasegawa}},
  \bibinfo{journal}{Nature Physics} \textbf{\bibinfo{volume}{8}},
  \bibinfo{pages}{185} (\bibinfo{year}{2012}).

\bibitem[{\citenamefont{Sulyok et~al.}(2013)\citenamefont{Sulyok, Sponar,
  Erhart, Badurek, Ozawa, and Hasegawa}}]{Sulyok13}
\bibinfo{author}{\bibfnamefont{G.}~\bibnamefont{Sulyok}},
  \bibinfo{author}{\bibfnamefont{S.}~\bibnamefont{Sponar}},
  \bibinfo{author}{\bibfnamefont{J.}~\bibnamefont{Erhart}},
  \bibinfo{author}{\bibfnamefont{G.}~\bibnamefont{Badurek}},
  \bibinfo{author}{\bibfnamefont{M.}~\bibnamefont{Ozawa}}, \bibnamefont{and}
  \bibinfo{author}{\bibfnamefont{Y.}~\bibnamefont{Hasegawa}},
  \bibinfo{journal}{Phys. Rev. A} \textbf{\bibinfo{volume}{88}},
  \bibinfo{pages}{022110} (\bibinfo{year}{2013}).

\bibitem[{\citenamefont{Demirel and Sponar}(2016)}]{Demirel16}
\bibinfo{author}{\bibfnamefont{B.}~\bibnamefont{Demirel}} 
\bibinfo{author}{\bibfnamefont{S.}~\bibnamefont{Sponar}} 
\bibinfo{author}{\bibfnamefont{G.}~\bibnamefont{Sulyok}} 
\bibinfo{author}{\bibfnamefont{M.}~\bibnamefont{Ozawa}} \bibnamefont{and}
  \bibinfo{author}{\bibfnamefont{Y.}~\bibnamefont{Hasegawa}},
  \bibinfo{journal}{Phys. Rev. Lett.} \textbf{\bibinfo{volume}{117}},
  \bibinfo{pages}{140402} (\bibinfo{year}{2016}).

\bibitem[{\citenamefont{Sulyok and Sponar}(2017)}]{Sponar17}
\bibinfo{author}{\bibfnamefont{G.}~\bibnamefont{Sulyok}} \bibnamefont{and}
  \bibinfo{author}{\bibfnamefont{S.}~\bibnamefont{Sponar}},
  \bibinfo{journal}{Phys. Rev. A} \textbf{\bibinfo{volume}{96}},
  \bibinfo{pages}{022137} (\bibinfo{year}{2017}).
  
  \bibitem[{\citenamefont{Demirel and Sponar}(2019)}]{Demirel19}
\bibinfo{author}{\bibfnamefont{B.}~\bibnamefont{Demirel}} 
\bibinfo{author}{\bibfnamefont{S.}~\bibnamefont{Sponar}} 
\bibinfo{author}{\bibfnamefont{A.A.}~\bibnamefont{Abbott}} 
\bibinfo{author}{\bibfnamefont{C.}~\bibnamefont{Branciard}} \bibnamefont{and}
  \bibinfo{author}{\bibfnamefont{Y.}~\bibnamefont{Hasegawa}},
  \bibinfo{journal}{New. J. Phys.} \textbf{\bibinfo{volume}{21}},
  \bibinfo{pages}{013038} (\bibinfo{year}{2019}).
  
  \bibitem[{\citenamefont{Demirel and Sponar}(2020)}]{Demirel20}
\bibinfo{author}{\bibfnamefont{B.}~\bibnamefont{Demirel}} 
\bibinfo{author}{\bibfnamefont{S.}~\bibnamefont{Sponar}}  \bibnamefont{and}
 \bibinfo{author}{\bibfnamefont{Y.}~\bibnamefont{Hasegawa}},
  \bibinfo{journal}{Appl. Sci.} \textbf{\bibinfo{volume}{10}},
  \bibinfo{pages}{1087} (\bibinfo{year}{2020}).
  
\bibitem[{\citenamefont{Rozema et~al.}(2012)\citenamefont{Rozema, Darabi,
  Mahler, Hayat, Soudagar, and Steinberg}}]{Steinberg12}
\bibinfo{author}{\bibfnamefont{L.~A.} \bibnamefont{Rozema}},
  \bibinfo{author}{\bibfnamefont{A.}~\bibnamefont{Darabi}},
  \bibinfo{author}{\bibfnamefont{D.~H.} \bibnamefont{Mahler}},
  \bibinfo{author}{\bibfnamefont{A.}~\bibnamefont{Hayat}},
  \bibinfo{author}{\bibfnamefont{Y.}~\bibnamefont{Soudagar}}, \bibnamefont{and}
  \bibinfo{author}{\bibfnamefont{A.~M.} \bibnamefont{Steinberg}},
  \bibinfo{journal}{Phys. Rev. Lett.} \textbf{\bibinfo{volume}{109}},
  \bibinfo{pages}{100404} (\bibinfo{year}{2012}).

\bibitem[{\citenamefont{Baek et~al.}(2013)\citenamefont{Baek, Kaneda, Ozawa,
  and Edamatsu}}]{Edamatsu13}
\bibinfo{author}{\bibfnamefont{S.-Y.} \bibnamefont{Baek}},
  \bibinfo{author}{\bibfnamefont{F.}~\bibnamefont{Kaneda}},
  \bibinfo{author}{\bibfnamefont{M.}~\bibnamefont{Ozawa}}, \bibnamefont{and}
  \bibinfo{author}{\bibfnamefont{K.}~\bibnamefont{Edamatsu}},
  \bibinfo{journal}{Scientific reports} \textbf{\bibinfo{volume}{3}},
  \bibinfo{pages}{2221} (\bibinfo{year}{2013}).

\bibitem[{\citenamefont{Kaneda et~al.}(2014)\citenamefont{Kaneda, Baek, Ozawa,
  and Edamatsu}}]{Kaneda14}
\bibinfo{author}{\bibfnamefont{F.}~\bibnamefont{Kaneda}},
  \bibinfo{author}{\bibfnamefont{S.-Y.} \bibnamefont{Baek}},
  \bibinfo{author}{\bibfnamefont{M.}~\bibnamefont{Ozawa}}, \bibnamefont{and}
  \bibinfo{author}{\bibfnamefont{K.}~\bibnamefont{Edamatsu}},
  \bibinfo{journal}{Phys. Rev. Lett.} \textbf{\bibinfo{volume}{112}},
  \bibinfo{pages}{020402} (\bibinfo{year}{2014}).

\bibitem[{\citenamefont{Ringbauer et~al.}(2014)\citenamefont{Ringbauer,
  Biggerstaff, Broome, Fedrizzi, Branciard, and White}}]{Ringbauer14}
\bibinfo{author}{\bibfnamefont{M.}~\bibnamefont{Ringbauer}},
  \bibinfo{author}{\bibfnamefont{D.~N.} \bibnamefont{Biggerstaff}},
  \bibinfo{author}{\bibfnamefont{M.~A.} \bibnamefont{Broome}},
  \bibinfo{author}{\bibfnamefont{A.}~\bibnamefont{Fedrizzi}},
  \bibinfo{author}{\bibfnamefont{C.}~\bibnamefont{Branciard}},
  \bibnamefont{and} \bibinfo{author}{\bibfnamefont{A.~G.} \bibnamefont{White}},
  \bibinfo{journal}{Phys. Rev. Lett.} \textbf{\bibinfo{volume}{112}},
  \bibinfo{pages}{020401} (\bibinfo{year}{2014}).

\bibitem[{\citenamefont{Ma et~al.}(2016)\citenamefont{Ma, Ma, Wang, Chen, Liu,
  Kong, Li, Peng, Shi, Shi et~al.}}]{Ma16}
\bibinfo{author}{\bibfnamefont{W.}~\bibnamefont{Ma}},
  \bibinfo{author}{\bibfnamefont{Z.}~\bibnamefont{Ma}},
  \bibinfo{author}{\bibfnamefont{H.}~\bibnamefont{Wang}},
  \bibinfo{author}{\bibfnamefont{Z.}~\bibnamefont{Chen}},
  \bibinfo{author}{\bibfnamefont{Y.}~\bibnamefont{Liu}},
  \bibinfo{author}{\bibfnamefont{F.}~\bibnamefont{Kong}},
  \bibinfo{author}{\bibfnamefont{Z.}~\bibnamefont{Li}},
  \bibinfo{author}{\bibfnamefont{X.}~\bibnamefont{Peng}},
  \bibinfo{author}{\bibfnamefont{M.}~\bibnamefont{Shi}},
  \bibinfo{author}{\bibfnamefont{F.}~\bibnamefont{Shi}}, \bibnamefont{et~al.},
  \bibinfo{journal}{Phys. Rev. Lett.} \textbf{\bibinfo{volume}{116}},
  \bibinfo{pages}{160405} (\bibinfo{year}{2016}).

\bibitem[{\citenamefont{Mao et~al.}(2019)\citenamefont{Mao, Ma, Jin, Sun, Fei,
  Zhang, Fan, and Pan}}]{Pan19}
\bibinfo{author}{\bibfnamefont{Y.-L.} \bibnamefont{Mao}},
  \bibinfo{author}{\bibfnamefont{Z.-H.} \bibnamefont{Ma}},
  \bibinfo{author}{\bibfnamefont{R.-B.} \bibnamefont{Jin}},
  \bibinfo{author}{\bibfnamefont{Q.-C.} \bibnamefont{Sun}},
  \bibinfo{author}{\bibfnamefont{S.-M.} \bibnamefont{Fei}},
  \bibinfo{author}{\bibfnamefont{Q.}~\bibnamefont{Zhang}},
  \bibinfo{author}{\bibfnamefont{J.}~\bibnamefont{Fan}}, \bibnamefont{and}
  \bibinfo{author}{\bibfnamefont{J.-W.} \bibnamefont{Pan}},
  \bibinfo{journal}{Phys. Rev. Lett.} \textbf{\bibinfo{volume}{122}},
  \bibinfo{pages}{090404} (\bibinfo{year}{2019}).

\bibitem[{\citenamefont{Ozawa}(2003{\natexlab{b}})}]{OzawaPLA03}
\bibinfo{author}{\bibfnamefont{M.}~\bibnamefont{Ozawa}},
  \bibinfo{journal}{Physics Letters A} \textbf{\bibinfo{volume}{318}},
  \bibinfo{pages}{21 } (\bibinfo{year}{2003}).

\bibitem[{\citenamefont{Hall}(2004)}]{Hall04}
\bibinfo{author}{\bibfnamefont{M.~J.~W.} \bibnamefont{Hall}},
  \bibinfo{journal}{Phys. Rev. A} \textbf{\bibinfo{volume}{69}},
  \bibinfo{pages}{052113} (\bibinfo{year}{2004}).

\bibitem[{\citenamefont{Busch et~al.}(2014{\natexlab{a}})\citenamefont{Busch,
  Lahti, and Werner}}]{Busch13PRA}
\bibinfo{author}{\bibfnamefont{P.}~\bibnamefont{Busch}},
  \bibinfo{author}{\bibfnamefont{P.}~\bibnamefont{Lahti}}, \bibnamefont{and}
  \bibinfo{author}{\bibfnamefont{R.~F.} \bibnamefont{Werner}},
  \bibinfo{journal}{Phys. Rev. A} \textbf{\bibinfo{volume}{89}},
  \bibinfo{pages}{012129} (\bibinfo{year}{2014}{\natexlab{a}}).

\bibitem[{\citenamefont{Busch et~al.}(2014{\natexlab{b}})\citenamefont{Busch,
  Lahti, and Werner}}]{Busch14}
\bibinfo{author}{\bibfnamefont{P.}~\bibnamefont{Busch}},
  \bibinfo{author}{\bibfnamefont{P.}~\bibnamefont{Lahti}}, \bibnamefont{and}
  \bibinfo{author}{\bibfnamefont{R.~F.} \bibnamefont{Werner}},
  \bibinfo{journal}{Rev. Mod. Phys.} \textbf{\bibinfo{volume}{86}},
  \bibinfo{pages}{1261} (\bibinfo{year}{2014}{\natexlab{b}}).

\bibitem[{\citenamefont{Buscemi et~al.}(2014)\citenamefont{Buscemi, Hall,
  Ozawa, and Wilde}}]{Buscemi14}
\bibinfo{author}{\bibfnamefont{F.}~\bibnamefont{Buscemi}},
  \bibinfo{author}{\bibfnamefont{M.~J.} \bibnamefont{Hall}},
  \bibinfo{author}{\bibfnamefont{M.}~\bibnamefont{Ozawa}}, \bibnamefont{and}
  \bibinfo{author}{\bibfnamefont{M.~M.} \bibnamefont{Wilde}},
  \bibinfo{journal}{Phys. Rev. Lett.} \textbf{\bibinfo{volume}{112}},
  \bibinfo{pages}{050401} (\bibinfo{year}{2014}).

\bibitem[{\citenamefont{Sulyok et~al.}(2015)\citenamefont{Sulyok, Sponar,
  Demirel, Buscemi, Hall, Ozawa, and Hasegawa}}]{Sulyok15}
\bibinfo{author}{\bibfnamefont{G.}~\bibnamefont{Sulyok}},
  \bibinfo{author}{\bibfnamefont{S.}~\bibnamefont{Sponar}},
  \bibinfo{author}{\bibfnamefont{B.}~\bibnamefont{Demirel}},
  \bibinfo{author}{\bibfnamefont{F.}~\bibnamefont{Buscemi}},
  \bibinfo{author}{\bibfnamefont{M.~J.~W.} \bibnamefont{Hall}},
  \bibinfo{author}{\bibfnamefont{M.}~\bibnamefont{Ozawa}}, \bibnamefont{and}
  \bibinfo{author}{\bibfnamefont{Y.}~\bibnamefont{Hasegawa}},
  \bibinfo{journal}{Phys. Rev. Lett.} \textbf{\bibinfo{volume}{115}},
  \bibinfo{pages}{030401} (\bibinfo{year}{2015}).

\bibitem[{\citenamefont{Barchielli et~al.}(2018)\citenamefont{Barchielli,
  Gregoratti, and Toigo}}]{Barchielli18}
\bibinfo{author}{\bibfnamefont{A.}~\bibnamefont{Barchielli}},
  \bibinfo{author}{\bibfnamefont{M.}~\bibnamefont{Gregoratti}},
  \bibnamefont{and} \bibinfo{author}{\bibfnamefont{A.}~\bibnamefont{Toigo}},
  \bibinfo{journal}{Communications in Mathematical Physics}
  \textbf{\bibinfo{volume}{357}}, \bibinfo{pages}{1253} (\bibinfo{year}{2018}).

\bibitem[{\citenamefont{Deutsch}(1983)}]{Deutsch83}
\bibinfo{author}{\bibfnamefont{D.}~\bibnamefont{Deutsch}},
  \bibinfo{journal}{Phys. Rev. Lett.} \textbf{\bibinfo{volume}{50}},
  \bibinfo{pages}{631} (\bibinfo{year}{1983}).

\bibitem[{\citenamefont{Kraus}(1987)}]{Kraus87}
\bibinfo{author}{\bibfnamefont{K.}~\bibnamefont{Kraus}},
  \bibinfo{journal}{Phys. Rev. D} \textbf{\bibinfo{volume}{35}},
  \bibinfo{pages}{3070} (\bibinfo{year}{1987}).

\bibitem[{\citenamefont{Maassen and Uffink}(1988)}]{Maassen88}
\bibinfo{author}{\bibfnamefont{H.}~\bibnamefont{Maassen}} \bibnamefont{and}
  \bibinfo{author}{\bibfnamefont{J.~B.~M.} \bibnamefont{Uffink}},
  \bibinfo{journal}{Phys. Rev. Lett.} \textbf{\bibinfo{volume}{60}},
  \bibinfo{pages}{1103} (\bibinfo{year}{1988}).

\bibitem[{\citenamefont{Berta et~al.}(2010)\citenamefont{Berta, Christandl,
  Colbeck, Renes, and Renner}}]{Berta2010}
\bibinfo{author}{\bibfnamefont{M.}~\bibnamefont{Berta}},
  \bibinfo{author}{\bibfnamefont{M.}~\bibnamefont{Christandl}},
  \bibinfo{author}{\bibfnamefont{R.}~\bibnamefont{Colbeck}},
  \bibinfo{author}{\bibfnamefont{J.~M.} \bibnamefont{Renes}}, \bibnamefont{and}
  \bibinfo{author}{\bibfnamefont{R.}~\bibnamefont{Renner}},
  \bibinfo{journal}{Nature Physics} \textbf{\bibinfo{volume}{6}},
  \bibinfo{pages}{659} (\bibinfo{year}{2010}).

\bibitem[{\citenamefont{Pati et~al.}(2012)\citenamefont{Pati, Wilde, Devi,
  Rajagopal, and Sudha}}]{Pati12}
\bibinfo{author}{\bibfnamefont{A.~K.} \bibnamefont{Pati}},
  \bibinfo{author}{\bibfnamefont{M.~M.} \bibnamefont{Wilde}},
  \bibinfo{author}{\bibfnamefont{A.~R.~U.} \bibnamefont{Devi}},
  \bibinfo{author}{\bibfnamefont{A.~K.} \bibnamefont{Rajagopal}},
  \bibnamefont{and} \bibinfo{author}{\bibnamefont{Sudha}},
  \bibinfo{journal}{Phys. Rev. A} \textbf{\bibinfo{volume}{86}},
  \bibinfo{pages}{042105} (\bibinfo{year}{2012}).

\bibitem[{\citenamefont{Abbott et~al.}(2016)\citenamefont{Abbott, Alzieu, Hall,
  and Branciard}}]{Abbott16}
\bibinfo{author}{\bibfnamefont{A.~A.} \bibnamefont{Abbott}},
  \bibinfo{author}{\bibfnamefont{P.-L.} \bibnamefont{Alzieu}},
  \bibinfo{author}{\bibfnamefont{M.~J.~W.} \bibnamefont{Hall}},
  \bibnamefont{and}
  \bibinfo{author}{\bibfnamefont{C.}~\bibnamefont{Branciard}},
  \bibinfo{journal}{Mathematics} \textbf{\bibinfo{volume}{4}}
  (\bibinfo{year}{2016}).

\end{thebibliography}
\end{document}